\documentclass[aps,prb,twocolumn,reprint,
			   groupedaddress,superscriptaddress,
			   amsfonts,amssymb,amsmath,a4paper]{revtex4-1}

\usepackage{graphicx,hyperref}

\usepackage{xcolor}
\usepackage{graphicx}
\usepackage{tikz}
\usepackage[shadow,textsize=small,textwidth=1.8cm,
backgroundcolor=green!20!white,
linecolor=black]{todonotes}
\usepackage{hyperref}
\hypersetup{
    colorlinks,
    linkcolor={blue!75!black!80!yellow},
    citecolor={blue!75!black!80!yellow},
    urlcolor={blue!75!black!80!yellow}
}

\usepackage[centering,hmargin=1.85cm,tmargin=31mm,bmargin=37mm]{geometry}

\graphicspath{{./graphicx/}}
\color{black!88!white}

\makeatletter
\renewcommand{\fnum@figure}{\figurename~\thefigure\ (color online)}

\makeatother

\begin{document}

\title{Localized Surface Plasmons in Vibrating Graphene Nanodisks}

\author{Weihua Wang}
\affiliation{Department of Photonics Engineering, Technical University of Denmark, DK-2800 Kgs. Lyngby, Denmark}
\affiliation{Center for Nanostructured Graphene, Technical University of Denmark, DK-2800 Kgs. Lyngby, Denmark}

\author{Bo-Hong Li}
\affiliation{Department of Photonics Engineering, Technical University of Denmark, DK-2800 Kgs. Lyngby, Denmark}
\affiliation{Center for Nanostructured Graphene, Technical University of Denmark, DK-2800 Kgs. Lyngby, Denmark}

\author{Erik Stassen}
\affiliation{Department of Photonics Engineering, Technical University of Denmark, DK-2800 Kgs. Lyngby, Denmark}
\affiliation{Center for Nanostructured Graphene, Technical University of Denmark, DK-2800 Kgs. Lyngby, Denmark}

\author{N.~Asger~Mortensen}
\email[]{asger@mailaps.org}
\affiliation{Department of Photonics Engineering, Technical University of Denmark, DK-2800 Kgs. Lyngby, Denmark}
\affiliation{Center for Nanostructured Graphene, Technical University of Denmark, DK-2800 Kgs. Lyngby, Denmark}

\author{Johan Christensen}
\email[]{jochri@fotonik.dtu.dk}
\affiliation{Department of Photonics Engineering, Technical University of Denmark, DK-2800 Kgs. Lyngby, Denmark}

\date{\today}

\begin{abstract}
Localized surface plasmons are confined collective oscillations of electrons in metallic nanoparticles. When driven by light, the optical response is dictated by geometrical parameters and the dielectric environment and plasmons are therefore extremely important for sensing applications. Plasmons in graphene disks have the additional benefit to be highly tunable via electrical stimulation. Mechanical vibrations create structural deformations in ways where the excitation of localized surface plasmons can be strongly modulated. We show that the spectral shift in such a scenario is determined by a complex interplay between the symmetry and shape of the modal vibrations and the plasmonic mode pattern. Tuning confined modes of light in graphene via acoustic excitations, paves new avenues in shaping the sensitivity of plasmonic detectors, and in the enhancement of the interaction with optical emitters, such as molecules, for future nanophotonic devices.
\end{abstract}

\maketitle

\section{Introduction}
To concentrate an optical field in nanometric volumes is challenging but nanophotonic advances teach us how to control and collect light in the form of surface plasmons at such small scale. Collective plasmon oscillations are confined at the interface between a metal and a dielectric and stem from the interaction of light and the conduction electrons in the metal. Localized surface plasmons (LSPs) on the other hand are coherent oscillations of the electric field generated when light waves are trapped within a conductive nanoparticle. The frequency of the electron oscillation does not only depend on the composition of the metal and its dielectric environment but the optical response is likewise determined by the particle size and shape. This paradigm gives rise to "plasmonic colouring effects" as has been used for centuries in stained glass, which is being revisited recently,\cite{Zhu:2015} but it has also laid the foundations to highly efficient devices for biochemical and medical testing given the pronounced sensitivity of the LSP resonances.\cite{Gramotnev:2010,Baev:2015}

With the birth of graphene plasmonics new routes have been enabled to gain an even more profound control for strong light-matter interactions. Graphene is known for its superior electrical,\cite{CastroNeto:2009,DasSarma:2011} mechanical,\cite{Bunch:2007,Lee:2008} thermal,\cite{Balandin:2011} and optical properties\cite{Nair:2008,Blake:2007} along with a potential for novel applications and technology.\cite{Ferrari:2015} Graphene plasmons hold many advantages as opposed to plasmons in noble metals since they are more strongly confined.\cite{Garcia-de-Abajo:2014} Graphene can be gated to constitute highly tunable and low-loss guides for light in future compact plasmon devices, which are useful for signal processing and infrared sensing.\cite{Jablan:2009,Koppens:2011,Christensen:2012,Chen:2012,Fei:2012} The linear dispersion of the Dirac fermions give rise to a broad landscape of tunable optical properties and collective plasmons oscillations. Hence, the implementation of graphene for tunable nanophotonic devices such as modulators, switches and photodetectors simply underpins the rapidly growing interest in this field.\cite{Liu:2011a,Liu:2011b,Lee:2012,Koppens:2014}

Apart from the geometrical layout in nanostructured graphene that determines the optical response, plasmons in curved and nonplanar graphene geometries are also being considered.\cite{Christensen:2015,Kumar:2014,Riso:2015,Smirnova:2015} Likewise, electrically doping the system with an applied electric field proves to be a versatile technique for tuning graphene nanodisk plasmons. Controlling LSPs and their mutual interactions via external stimulus, enabled control of the enhancement of the photon absorption with application to infrared opto-electric devices as recent experiments demonstrated.\cite{Fang:2014} Rather than tuning the optical response via an electric field or chemical doping, current simulations show that mechanical vibrations in extended unpatterned graphene sheets create the necessary momentum to efficiently couple light into plasmons.\cite{Farhat:2013,Schiefele:2013} These findings suggest that new plasmonic devices can be designed whose spectral sensitivity, photoresponse, photon absorption and modulation speed could be tailored by sound.\cite{Yan:2012,Freitag:2013}

In this paper, we add a new ingredient to actively control localized graphene plasmons through mechanical deformations.\cite{Bunch:2008,Boddeti:2013,Midtvedt:2014} Instead of changing the properties of LSPs through varying the charge carrier concentration, deformation of graphene disks through vibrations produces observable plasmon shifts. Further to this, we show that the acoustic modes can strongly modify the decay rates of nearby emitters,\cite{Reserbat-Plantey:2015} such as a quantum dot or excited molecules, when placed in the nearest proximity of a graphene disk as illustrated in Fig.~1. In this context it is predicted that the modal shape of the vibrating nanodisk determines the unprecedented degree of the light-matter interaction enhancement when stimulated by sound. The fabrication of the resonator could be realized by depositing a graphene disk on a perforated $\text{SiO}_2$ layer and clamping it to the edges by attaching a thin insulation layer with a hole on top of it. The acoustical excitation of graphene nanodisks can be practically realized by a focused laser with frequency modulated intensity. Pulsing the laser generates phototermal excitations that create mechanical stress and cause the graphene nanodisks to vibrate at the modulation frequency.

\section{Mechanical vibrations of graphene nanodisks}
A monolayer graphene nanodisk that is clamped at the outer rim can commonly be described as a pre-tensioned vibrating circular membrane.\cite{Atalaya:2008,Chen:2009} The governing equation for time harmonic undamped transverse vibrations $\Psi(\bm r, t)$ (along the $z$ direction) reads
\begin{equation}
v^2\nabla^2\Psi(\bm r, t)+\Omega^2\Psi(\bm r, t)=-\frac{f(\Omega)}{\rho_sS},
\end{equation}
where $f(\Omega)$ is the external force stimulus of frequency $\Omega$ and, $S$ the surface area of the nanodisk. The speed of sound $v=\sqrt{T/\rho_s}$ is very sensitive to the membrane preparation and can be altered by pre-stressing the nanodisk with a tension $T$ by assuming an unaltered surface mass density $\rho_s$. In our investigation we choose an average tension of $T=0.2$ N $\text{m}^{-1}$ and the surface mass density $\eta_s=2.96\times10^{-6}$ kg $\text{m}^{-2}$ as reported in the literature.\cite{Lee:2012} In order to find the mechanical eigenmodes we seek the homogeneous solution to an undriven problem. Hence, with fixed edges this solution reads
\begin{equation}
\Psi_{mn}(r, t)=A_{mn}J_{m}(k_{mn}r)\cos(m\theta)\sin(\Omega_{mn} t),
\end{equation}
where $A_{mn}$ is the complex modal amplitude, $J_m$ the Bessel function of the first kind of order $m$, and $k_{mn}=j_{mn}/R$ where $j_{mn}$ is the $n$th root of $J_m$. From this expression we get a linear relationship between frequency and disk radius, $\Omega_{mn}=j_{mn}v/R$.

\section{Optical excitations of graphene nanodisks}
The optical response of graphene structures relies on treating the material as infinitely thin with a finite conductivity. For the sufficiently large diameters we can safely ignore atomic details\cite{Christensen:2014,Wang:2015} and nonlocal effects.\cite{Mortensen:2014,Wang:2011,Wang:2012} In the electrostatic limit we seek the solutions of Maxwell's equations (with time dependence $e^{-i\omega t}$) through a self-consistent relation:\cite{Wang:2011,Wang:2012}
\begin{equation}
\begin{split}
\int_{S}d\bm r&\Phi(\bm r)\rho(\bm r)= \frac{\sigma(\omega)}{i\omega}\int_{S}d\bm r\nabla\Phi(\bm r)\cdot\nabla\Phi_{ext}(\bm r) \\
&+\frac{1}{4\pi\varepsilon_m}\frac{\sigma(\omega)}{i\omega}\int_{S}d\bm r\nabla\Phi(\bm r)\cdot\nabla\int_{S} d\bm r^\prime\frac{\rho(\bm r^\prime)}{|\bm r- \bm r^\prime|},
\end{split}
\end{equation}
representing conservation for both charge and energy. More specifically, $\rho(\bm r)$ is the induced surface charge density, $\Phi(\bm r)$ and $\Phi_{ext}(\bm r)$ are the total and external potentials, $\sigma(\omega)$ is the surface conductivity and, $\varepsilon_m$ is the averaged permitivity of the surrounding medium. For arbitrarily shaped graphene structures, Eq.~(3) can be solved numerically as we are going to clarify in more detail below.

\section{Modulation of LSP resonances by sound}
The electro-mechanical dynamics is a complex process, but thanks to the huge temporal mismatch between the interacting resonances, we are able to simplify the problem. The wave oscillation of the LSPs (at THz) occur at much faster pace compared to the mechanical modes (at GHz), which implies that the plasmons will follow the moving membrane adiabatically. Thus, at a given time plasmons only notice a static deformation rather than a rapidly vibrating body.

In practical terms, this means that we can initiate the numerical simulations by calculating the displacements of the acoustic modes, from which we get a static deformation of the graphene nanodisks that is needed to subsequently compute the electromagnetic (EM) response. In calculating the plasmonic properties, we solve Eq (3) through finite-elements, which concretely means that the potentials and induced charge densities are discretized over an acoustically deformed domain.\cite{Wang:2015} In all simulations, we choose a Fermi level $E_F=0.4$\,eV (resembling typical doping levels achieved in experiments\cite{Fang:2013,Fang:2014,Zhu:2014}), and $\varepsilon_m=1.35\varepsilon_0$ ($\varepsilon_0$ vacuum permittivity) accounting for the screening effect from the surroundings but without any phonon contributions.\cite{Zhu:2014}

We begin with the assumption that the disk is excited by a low intensity pulse such that the maximal disk displacement does not exceed the disk radius itself. We work in a spectral range where three different modes can be excited; $\Psi_{01}$, $\Psi_{02}$, and $\Psi_{11}$. Fixing the disk radius at $R=50$\,nm, these three modes vibrate around $\Omega_{mn}/2\pi=$ 2.0 GHz, 4.6 GHz, and 3.2 GHz respectively, corresponding to the following roots $j_{01}\simeq2.4$, $j_{02}\simeq5.5$, and $j_{11}\simeq3.8$. With these mechanical resonance frequencies, we are on the safe side of our assumptions since the plasmon frequency of the nanodisks without an acoustical load is as high as 24 THz. In Fig.~2(a) we excite the fundamental axisymmetric (0,1) mode with an amplitude of 50\,nm. Since the mechanical cycle determines the EM interaction, we compute the plasmon frequency over the mechanical phase upto a quarter period $\Omega t=\pi/2$. The elastostatic limit is met when $\Omega t\rightarrow 0$, meaning that the disks is not in motion and conventional LSPs in graphene are excited. As seen in Fig.~2(a), we inspect solely the lowest breathing and dipole modes since higher ordered resonances are very similar. When the mechanical phase accumulates, both resonances redshift but the breathing mode exhibits a more pronounced sensitivity as compared to the dipole resonance as depicted in Fig.~2a. The higher sensitivity results in a more notable shift. Since the plasmonic response is modified by sound, the plasmo-mechanic coupling can be appreciated from the density of states (DOS) $\rho(\omega)$ of the plasmonic nanodisk while considering the acoustic load introduced by the mechanical vibrations. Note from Fig.~2a that $\rho(\omega)$ peaks around the band edges (temporal \emph{van Hove} singularities), but also remains high within the band during the mechanical cycle.

Next, we investigate the $\Psi_{02}$ modal vibration, which again is axisymmetric but contain $m=2$ nodal circles as illustrated by the inset of Fig.~2(b). As compared to the fundamental $(0,1)$ resonance, the plasmonic breathing mode in the present example shows to exhibit a higher sensitive and is redshifted by almost 0.1 eV upto the point where it degenerates with the dipole mode. We will try to shed some light on the spectral shift owing to the mechanical perturbation. With a relatively small displacement field $\Psi(\bm r, t)$, we perform a Taylor expansion of $1/|\bm r-\bm r^\prime+\Psi(r, t)-\Psi(r^\prime, t)|$ and can qualitatively give the energy shift $\delta(\hbar\omega)$ of the LSPs induced by the electro-mechanical coupling as follows
\begin{equation}
\delta(\hbar\omega)\propto\int_{S}d\bm r\nabla\Phi(\bm r)\cdot\int_{S} d\bm r^\prime\frac{\rho(\bm r^\prime)\beta}{|\bm r- \bm r^\prime|^2}\nabla \Psi(r, t).
\end{equation}
In Eq.~(4), $\bm r$ and $\bm r^\prime $ run along the $x-y$ plane, and $\beta = |\Psi(r, t)-\Psi(r^\prime, t)|/|\bm r-\bm r^\prime|$. The axiymmetric displacements of mode $(0,1)$ and $(0,2)$ are seen to have only little influence on the plasmonic dipole mode. In these cases, the potential distributions and the induced charges, as visualized in the insets of Fig.~2, are mostly localized at the nanodisk edge when excited at the dipolar resonance. The EM energy is thus, predominately concentrated along the mechanically stationary rim ($|r|\rightarrow R$ and $|r^\prime|\rightarrow R$) where the numerator of $\beta$ goes to zero, causing the integral in Eq.~(4) to be very small. In this context, as expected, the energy variation $\delta(\hbar\omega)$ is more pronounced for the breathing mode. The charge distributes now throughout the bulk of the disks, which causes more efficient electro-mechanical coupling with more pronounced plasmonic sensitivity as measured by a large redshift. The order of the axisymmetric modes also play a significant role. The state $\Psi_{02}$ has two nodal circles and $\Psi_{01}$ only one. The distance between positive and negative charges along $z$ is thus enlarged for the $(0,2)$ mode due to the disk displacement and gives rise to a reduced Coulumb interaction and an increased redshift.

We finally study the mechanical mode of the $\Psi_{11}$ state, which is the lowest non-axisymmetric vibration with the radial nodal line along the $y$ axis as shown in the inset of Fig.~2(c). Interestingly, as we will see in the following, there is a certain interplay between this mechanical polarization and the orientation of the dipolar mode. While the breathing modes behaves similar to the mechanical $(0,1)$ mode from Fig.~2(a), two orthogonal dipolar modes do now exist that are degenerate in the static limit $\Omega t\rightarrow 0$. As the mechanical cycle sets in as depicted in Fig.~2(c), the dipolar mode along $x$ exhibits a redshift while the $y-$dipole mode blueshifts. The reduction of the Coulomb interaction in this case is also responsible for the redshift of the $x-$dipolar mode. The LSPs oscillate along the radial nodal line when the $y-$dipolar mode is excited. In this situation, as opposed to the $x$-dipolar mode, the mechanical vibration does not increase the separation of opposite charges, but instead, it increases the Coulomb interaction so the energy increase causes the LSPs to blueshift. By changing the frequency of the intensity modulated laser, we are able to excite many other mechanical modes, but we do not anticipate pronounced differences as compared to the previous discussion, which demonstrated an unique way to tune LSP resonances in graphene by acoustic stimulation.

\section{Tunable spontaneous emission}
Modifying the photonic environment of emitters to enhance the spontaneous emission, has been realized by means of both dielectric structures\cite{Englund:2005,Bleuse:2011} and plasmonic cavities,\cite{Vesseur:2010,Akselrod:2014,Tielrooij:2015} in order to gain full control of the light-matter interaction. In what follows, we propose a novel strategy by imposing acoustic wave excitation to dictate the efficiency of the spontaneous emission enhancement without altering the geometrical surroundings. As we show in Fig.~1, a dipole emitter is placed in the center at $z=60$ nm in front of the nanodisk. Again, we concentrate on a set of specifically mechanically excited states and study their temporal evolution through the accumulated phase, as illustrated in Fig.~3. We begin by recalling the enhancement of the spontaneous emission decay rate,\cite{Novotny:2006}
\begin{equation}
\frac{\Gamma}{\Gamma_0}=1+\frac{6\pi\varepsilon_0}{k^3}\frac{\text{Im}\{\bm\mu^*\cdot\bm E_{s}(\bm r_0)\}}{|\bm\mu|^2},
\end{equation}
where $k$ is the wave vector, $\bm\mu$ the dipole moment, and $\bm E_s(\bm r_0)$ the scattering field at the dipole's origin $\bm r_0$. The numerical results shown in Fig.~3 illustrate the complex interplay between the enhancement of the decay rate for specific dipole orientations and vibrational states. The peak locations of the Purcell factors are consistent with the eigenmodes calculated in Fig.~2, meaning that the resonances can be traced back to dipolar and breathing mode excitations of the nanodisks, when they are excited by an $x-$polarized and $z-$polarized emitter, respectively. Overall, we can see that the enhancement is remarkable with peak values approaching $10^4$.

In the following we will discuss on the selective enhancement process with respect to orientation of the dipole and the mechanical cycle. The $(0,1)$ mode makes the graphene disk move close to the emitter when the mechanical cycle $\Omega t /\pi$ advances. As seen in Figs.~3(a) and 3(b), this is true for emitters, both polarized along the $x$ and the $z$ axis. At a higher acoustic excitation frequency, the enhancement of the decay rate for a $z-$polarized emitter via the $\Psi_{02}$ modal displacement, as seen in Fig.~3(d), can be traced back to the same mechanism as before. If however, a dipole resonance is excited within the nanodisk where the electric charge confinement at the edge [see inset in Fig.~2(b)] moves away from the emitter due to the $(0,2)$ modal shape, the Purcell factor decreases with time as illustrated in Fig.~3(c). The mechanical vibration in the $\Psi_{11}$ state breaks the axial symmetry and determines the optical response with dependence of the in-plane dipole orientation. For $x-$polarized emitters, the excited dipole mode is effectively tilted by the mechanical displacement $\Psi_{11}$. This effect, causes the edge of the nanodisk actually to move closer to the emitter with time and induces an enhancement of the decay rate as shown in Fig.~3(e). When polarized parallel to radial nodal line, the induced dipole remains stationary in position, so the only effect is a slight blueshift as seen in Fig.~3(f). Finally, a complex interplay is predicted for the $z-$polarized case, giving rise to both Purcell factor enhancement and the excitation of x-polarized dipolar (the dotted elliptical regime) and breathing (the solid elliptical regime) resonances, see Fig.~3(g).

\section{Quantum calculations}

So far, our calculations of plasmons have been based on classical electrodynamics, Eq.~(3), and membrane displacement entered only the plasmon dynamics through the modified Coulomb interactions associated with the non-planar confinement of the electron gas. Here, we extend our study to include quantum considerations taking atomic-scale details and quantum plasmonics into account.\cite{Wang:2015} We will consider the small-displacement limit, where the curved-space corrections to the Coulomb interaction can be neglected, while strain effects on the electron states are still accounted for. We start from a tight-binding (TB) representation of the flat membrane to first calculate electronic states for the equilibrium case, i.e. in the absence of electromagnetic fields. Subsequently, we invoke the random-phase approximation (RPA) to calculate the energy-loss function associate with plasmon dynamics.\cite{Wang:2015} For membrane dynamics we fill focus our attention on sufficiently small deformations where the in-plane lattice strains are captured by an empirical tight-binding parameter parametrizations,\cite{Pereira:2009} that work fairly accurately for a relative bond-stretching less than 20\%.
In this model, the local tightbinding parameter $\gamma$ reflects the local change in bond lengths $a$ and follows the empirical expression
\cite{Pereira:2009}
\begin{equation}
\gamma = \gamma_0
\exp\left(−\frac{3.37︎a}{a_0}\,-\,1︎\right)
\end{equation}
with $a_0 = 1.42$\,\AA being the lattice constant of the unstrained lattice, while $\gamma_0=2.8$\,eV.\cite{Thongrattanasiri:2012}

For the purpose of illustration, we consider a disk of hexagonal shape with perfect six-fold rotational symmetry and an armchair edge termination. As a consequence, the lattice supports no additional electronic edgestates at the Fermi level\cite{Christensen:2014} and in accordance with group theory, both electronic states and plasmon states will be either nondegenerate or come has degenerate pairs.\cite{Wang:2015}

The hexagonal graphene disk that contains 366 carbon atoms, being equivalent to a sidelength of roughly 2\,nm. For the deflection amplitude $A_{nm}$ we limit ourselves to
0.4\,nm, which corresponds to no carbon-carbond bonds being stretched more
than 20\%.

For our RPA calculations at room temperature ($k_BT = 25.7$\,meV) we follow previous work\cite{Thongrattanasiri:2012,Wang:2015} and use a plasmon damping of $\hbar/\tau=1.6$\,meV and doping of the graphene disk corresponding to a Fermi energy of $0.4$\,eV.

In Fig.~4 we summarize the results which overall resembles the dynamics seen also from our classical electrodynamics considerations in Fig.~2. In particular, the plasmon breathing mode around 0.7\,eV is significantly modulated by the mechanical displacement of the membrane hosting the plasmons. In the example, none of the carbon-carbon bonds are stretched by more than 20\% (which is an upper critical limit for elastic deformation\cite{Geim:2009}), yet the plasmon energy modulation is as high as $100$\,meV corresponding to a relative modulation of up to 40\% for the dipole resonance while it is around 14\% for the breathing mode. This suggest that plasmon-energy modulation can be appreciable.

\section{Discussion and conclusions}

In this paper we have investigated the modulation of LSPs in graphene nanodisks by mechanical vibrations, which is a new route to tune graphene plasmons.

We note that classical electrodynamics have been used to illustrate the basic idea and for the ease of illustration, the membrane displacement is slightly exaggerated. However, even though our classical considerations will have their shortcomings, our full quantum treatment of plasmons in a realistically strained tight-binding lattice confirm the existence of mechanical modulation of graphene plasmons.

Rather than changing the shape and size of the disks or injecting charge carriers via doping or gating, appropriately chosen acoustic stimulations with respect to the modal profile, determines the very nature of the plasmonic response. This novel acousto-plasmonic system opens many striking possibilities to mold the flow of light at the nanoscale and to enhance light-matter interactions. We have speculated how this can be used to modulated the light-matter interactions, such as the decay dynamics of a quantum emitter in the near vicinity of the membrane. In fact, such an exciting experiments has in the meantime been done.\cite{Reserbat-Plantey:2015}

\section*{Acknowledgments}
We thank W. Yan, M. Wubs, and A.-P. Jauho for stimulating discussions. Center for Nanostructured Graphene is sponsored by Danish National Research Foundation, Project DNRF58. B.-H.L. and N.A.M. acknowledges the Villium Foundation (341/300-123012) and J.C. and N.A.M. acknowledges Danish Council for Independent Research (FTP 12-134776 and FNU 1323-00087). We thank F.~Koppens for private communications in relation to Ref.~\onlinecite{Reserbat-Plantey:2015}.

\newpage


%

\newpage

\begin{figure}[b!]
\includegraphics[width=7.0cm]{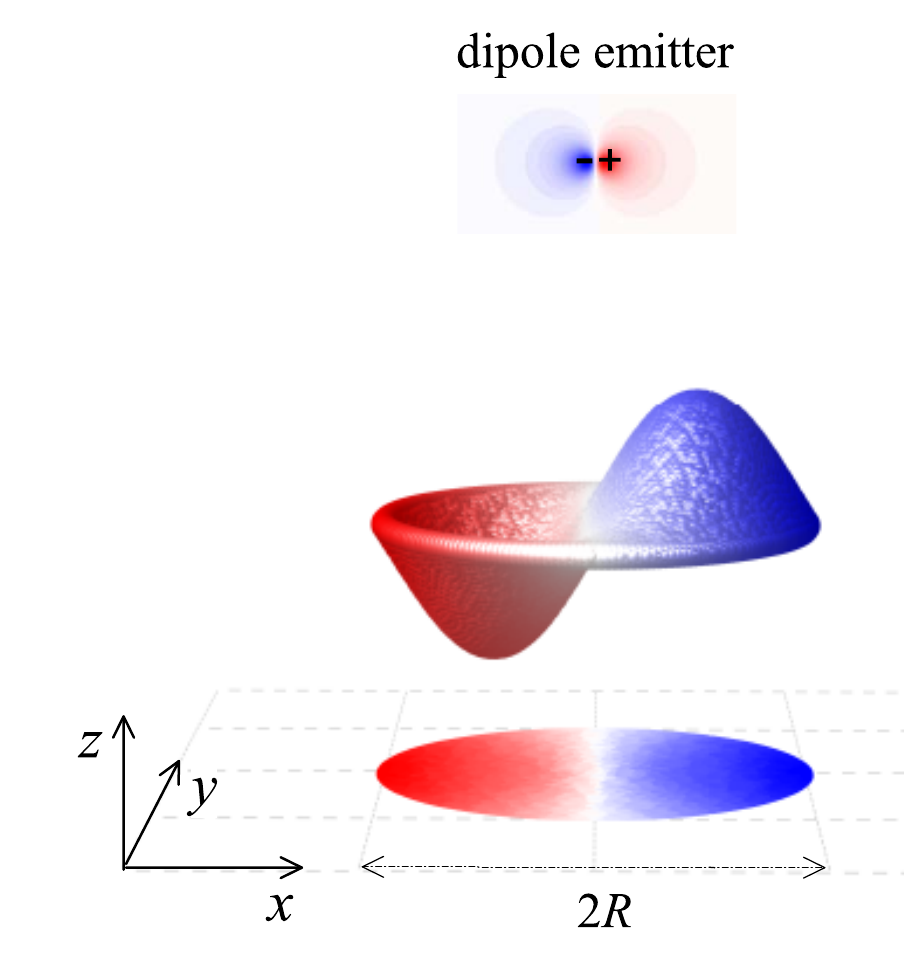}
\caption{Schematic diagram of the system under investigation: A graphene nanodisk with radius $R$ vibrates at its lowest asymmetric mode $\Psi_{11}$. An illustrative example is depicted showing a quantum dot simulated as a x-polarized dipole emitter situated nearby the nanodisk. We compute the induced near-field potential as excited by the emitter.}
\end{figure}

\begin{figure*}[]
\begin{center}
\includegraphics[width=18.0cm]{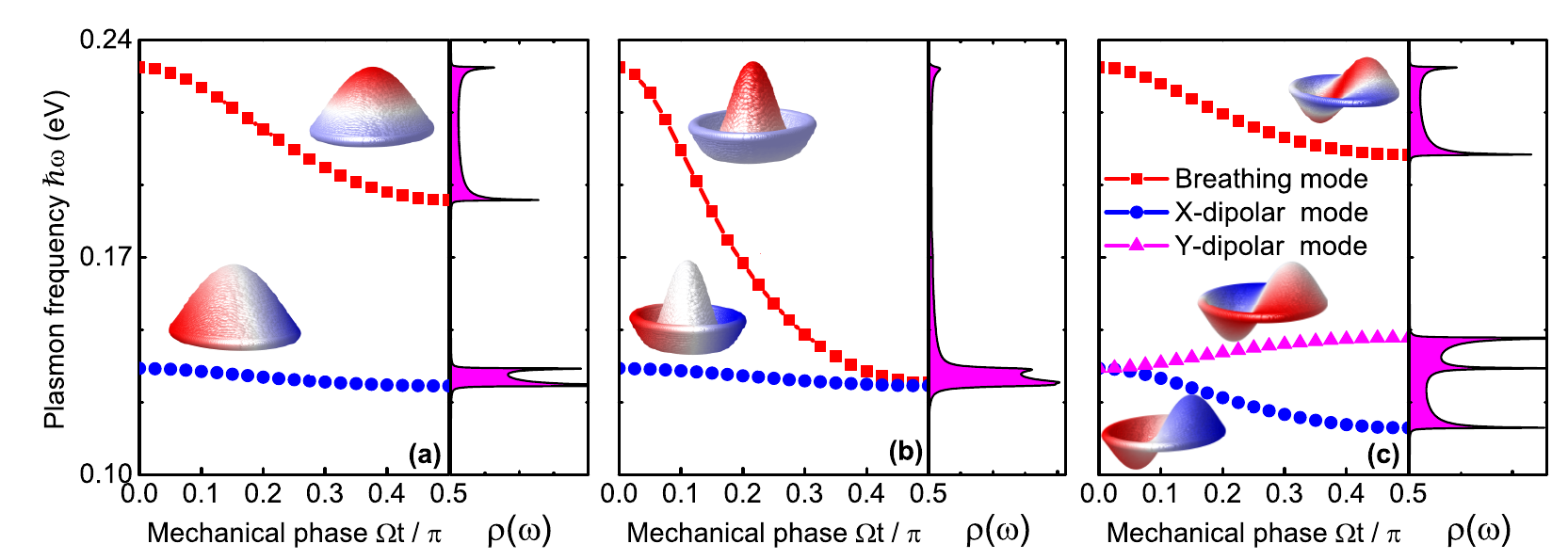}
\caption{Plasmon frequency as a function of the mechanical phase $\Omega t/\pi$ for (a) $\Psi_{01}$, (b) $\Psi_{02}$, and (c) $\Psi_{11}$. The spectra show only the lowest energy plasmon resonances, which are the breathing (squares) and dipole modes (circles and triangles). The insets show the displacements $\Psi_{mn}$ in their respective mechanical states together with the potential distributions for the specific plasmon modes.}
\end{center}
\end{figure*}

\begin{figure}[b!]
\includegraphics[width=8.0cm]{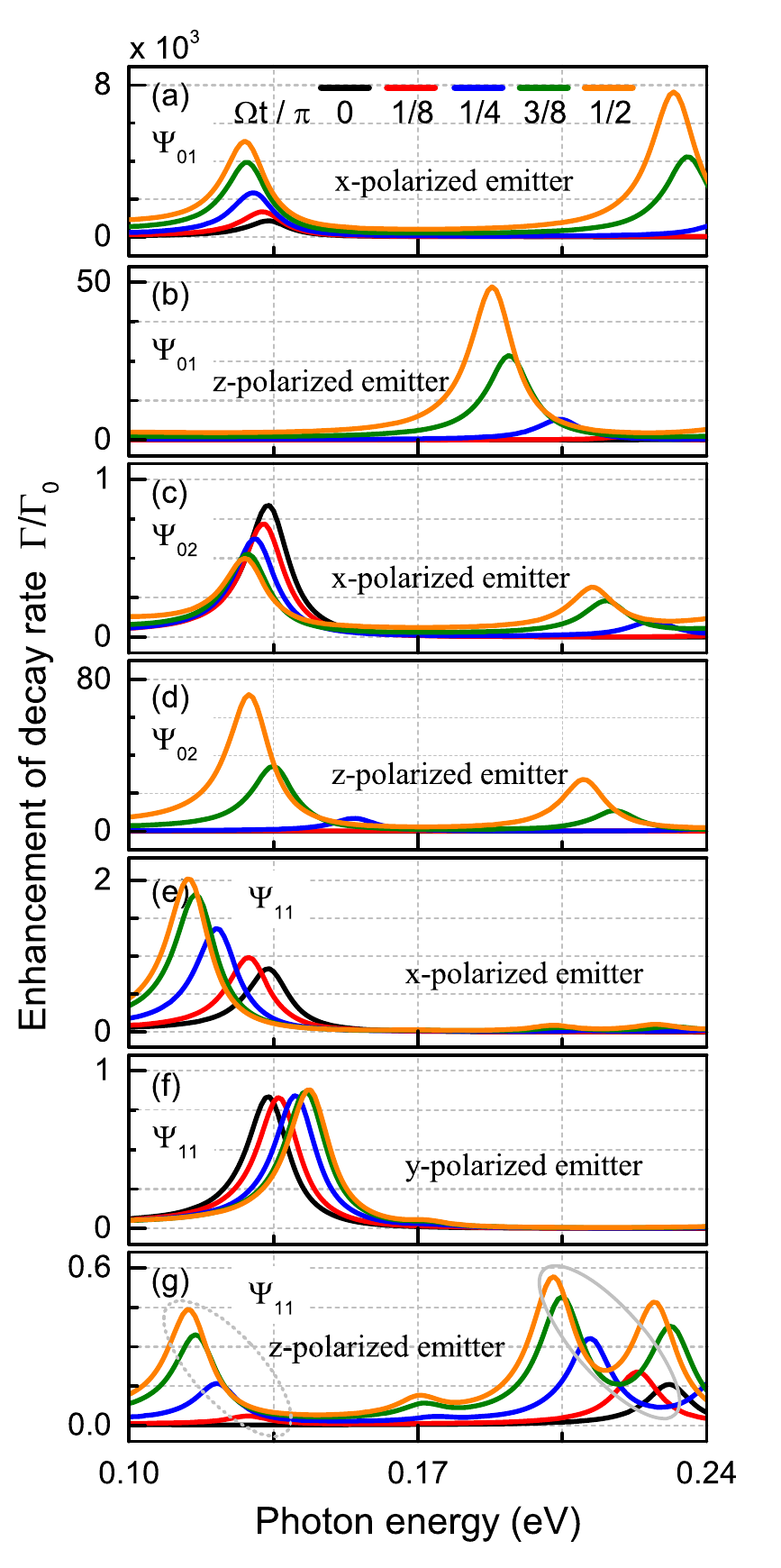}
\caption{Purcell factor for dipole emitters with $x-$, $y-$, and $z-$polarizations placed 60 nm away from the center ($z=0$) of the vibrating nanodisk. The emission decay rate $\Gamma$ is normalized to the vacuum rate $\Gamma_0$, and plotted against the photon emission energy at different values of the mechanical phase $\Omega t /\pi$. We consider the following displacement fields: (a,b) $\Psi_{01}$, (c,d) $\Psi_{02}$, and (e-g) $\Psi_{11}$. The various dipole polarizations are indicated in the illustration.}
\end{figure}

\begin{figure}[t]
\includegraphics[width=\columnwidth]{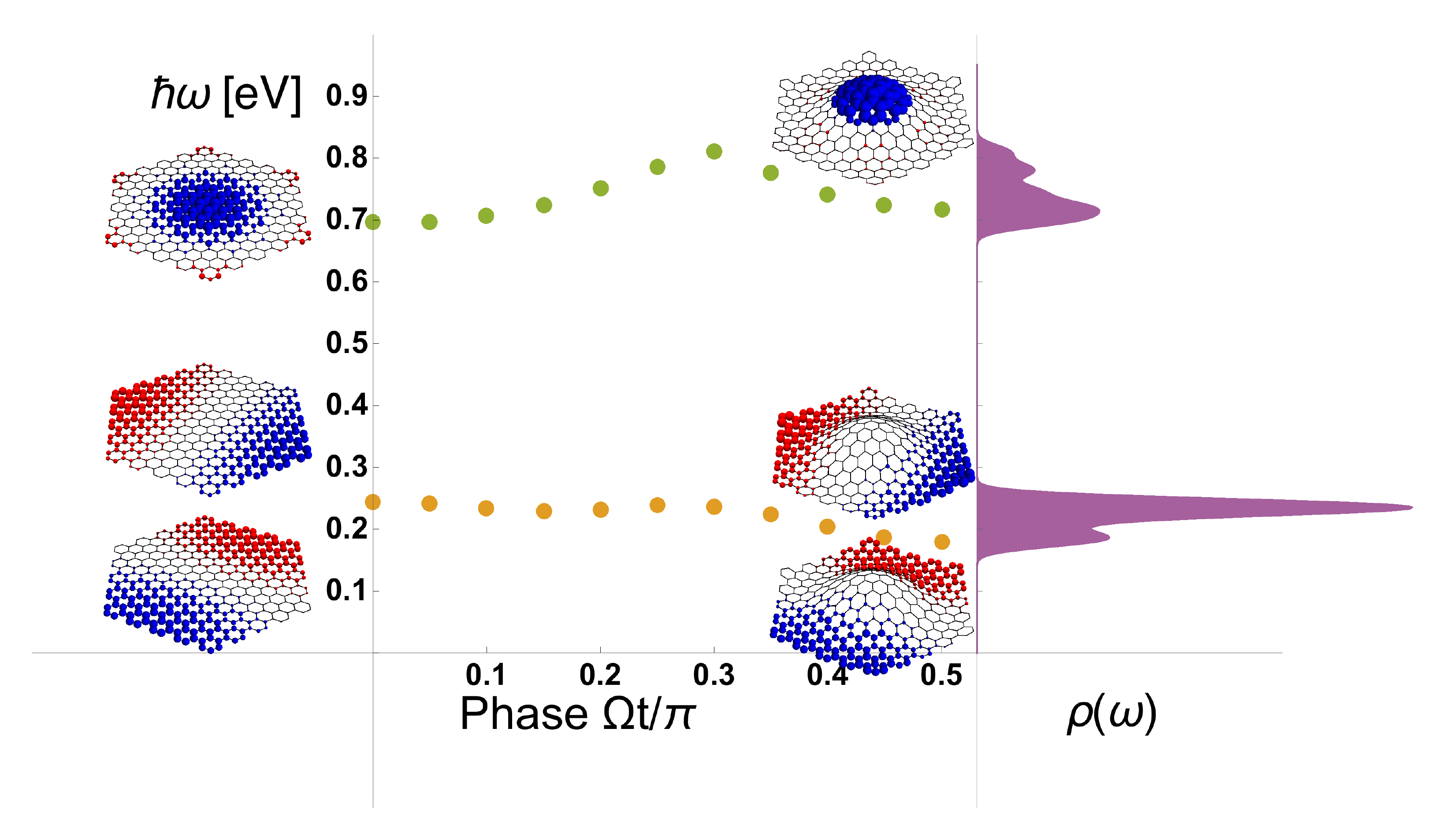}
\caption{TB-RPA calculations of quantum plasmons in a 366 atoms graphene membrane with a mechanical $\Psi_{02}$ displacement mode. Details of the parameters are discussed in the text. Note how the low-energy dipole modes are degenerate, while the high-energy breathing mode is nondegenerate.}
\end{figure}

\end{document}